\documentclass[DVI]{cej}
\usepackage{layout}
\usepackage{amsmath}
\usepackage{textcomp}
\usepackage{mathrsfs}
\usepackage{amsfonts}
\usepackage{amssymb}
\newcommand\mr{\mathscr}

\newcommand\mb{\mathbb}

\def\m{\mu}
\def\n{\nu}

\def\p{\partial}

\def\lag{\langle}
\def\rag{\rangle}

\title{The chiral and deconfinement phase transitions}

\articletype{Research Article}

\author{Fukun Xu \inst{1}\email{xufukun@mail.ihep.ac.cn},
            Mei Huang \inst{1}$^,$\inst{2}\email{huangm@mail.ihep.ac.cn}}

\institute{
     \inst{1} Institute of High Energy Physics, Chinese Academy of Sciences, \\
     Yuquan Road 19B, 100049, Beijing, China
     \inst{2} Theoretical Physics Center for Science Facilities, Chinese Academy of Sciences, \\
    Yuquan Road 19B, 100049, Beijing, China
          }

\abstract{By introducing the dressed Polayakov loop or dual chiral condensate as a candidate order
parameter to describe the deconfinement phase transition for light flavors,  we discuss the interplay
between the chiral and deconfinement phase transitions, and propose the possible QCD phase diagram
at finite temperature and density. We also introduce a dynamical gluodynamic model with
dimension-2 gluon condensate, which can describe the color electric deconfinement as well as
color magnetic confinement. }


\begin{document}
\maketitle


\section{Introduction}

The interplay between chiral and deconfinement phase transitions at
finite temperature and density are of continuous interests for
studying the QCD phase diagram. The chiral restoration is characterized
by the restoration of chiral symmetry and the deconfinement phase transition
is characterized by the breaking of center symmetry, which are only well defined
in two extreme quark mass limits, respectively. In the chiral limit when
the current quark mass is zero $m=0$, the chiral condensate
$\langle{\bar q}q \rangle$ is the order parameter for the chiral
phase transition. When the current quark mass goes to infinity
$m\rightarrow \infty$, QCD becomes pure gauge $SU(3)$ theory, which
is center symmetric in the vacuum, and the usually used order
parameter is the Polyakov loop expectation value $\langle P \rangle
$ \cite{Polyakov:1978vu}, which is related to the heavy quark free
energy.

At zero density and chiral limit, lattice QCD results show
that the chiral and deconfinement phase transitions occur at the
same critical temperature \cite{Kogut:1982rt}.
It has been largely believed for a long time that chiral symmetry
restoration always coincides with deconfinement phase transition in
the whole $(T,\mu)$ plane. It has been conjectured in Ref. \cite{McLerran:2007qj}
that in large $N_c$ limit, a confined but chiral symmetric phase,
which is called quarkyonic phase can exist in the high baryon
density region. It attracts a lot of interests to study whether this
quarkyonic phase can survive in real QCD phase diagram.

For the case of finite physical quark mass, neither the chiral condensate
nor the Polyakov loop is a good order parameter. The Wuppetal-Budapest group \cite{WB}
found that for the case of $N_f=2+1$, there are three pseudocritical
temperatures, the transition temperature for chiral restoration of
$u,d$ quarks $T_c^{\chi(ud)}= 151(3)(3)~{\rm MeV}$, the transition
temperature for $s$ quark number susceptibility
$T_c^{s}=175(2)(4){\rm MeV}$ and the deconfinement transition
temperature $T_c^d=176(3)(4) {\rm MeV}$ from the Polyakov loop.
These results are agreed by the hotQCD collaboration  \cite{HotQCD} by
using an improved HISQ action.  (The recent extracted
critical temperature for deconfinement phase transition from RHIC data is
$T_c^{d}=175_{-7}^{+1} {\rm MeV}$ \cite{Gupta:2011wh}. )

This talk aims to discuss two topics related to deconfinement phase transition.
In Sec. \ref{sec-chiral-deconf}, we will discuss the possible order parameter candidate for describing
the deconfinement phase transition for light flavors, and investigate the interplay
between the chiral and deconfinement phase transitions, based on our results,
we will propose the possible QCD phase diagram at finite temperature and density.
In Sec. \ref{sec-gluodynamic}, we will try to introduce a gluodynamic model to describe
the deconfinement phase transition.

\section{Dressed Polyakov loop and deconfinement phase transition for light flavors}
\label{sec-chiral-deconf}

Recent investigation revealed that quark propagator, heat kernels
can also act as an order parameter as they transform non trivially
under the center transformation related to deconfinement transition
\cite{Gattringer:2006ci,Synatschke,Gattringer}. The
exciting result is the behavior of spectral sum of the Dirac
operator under center transformation.
A new order parameter, called dressed Polyakov loop has been defined
which can be represented as a spectral sum of the Dirac operator. It has
been found the infrared part of the spectrum particularly plays a leading role
in confinement. This result is encouraging since it gives
a hope to relate the chiral phase transition with the
confinement-deconfinement phase transition. The order parameter for
chiral phase transition is related to the spectral density of the
Dirac operator through Banks-Casher relation \cite{Banks:1979yr}.
Therefore, both the dressed Polyakov loop and the chiral condensate
are related to the spectral sum of the Dirac operator.

Consider a $U(1)$ valued boundary condition for the fermionic
fields in the temporal direction
$\psi(x,\beta) = e^{-i \phi} \psi(x,0)$,
where $0\leq \phi < 2\pi $ is the phase angle and $\beta$ is the
inverse temperature. The dual quark condensate or the dressed Polyakov loop $\Sigma_1$ is then defined
as $\Sigma_1 =-{\int_0}^{2\pi} \frac{d\phi}{2\pi} e^{-i \phi}
\langle\bar{\psi} \psi\rangle_\phi$. It transforms in the same way as the
conventional thin Polyakov loop
under the center symmetry and hence is an order parameter for the
deconfinement transition. It reduces to
the thin Polyakov loop and to the dual of the conventional chiral
condensate in infinite and zero quark mass limits respectively,
i.e., in the chiral limit $m \rightarrow 0$ we get the dual of the
conventional chiral condensate and in the $m\rightarrow \infty$
limit we have thin Polyakov loop. Therefore, we extend the dressed Polyakov
loop as a candidate order parameter to describe the deconfinement phase transition
of a quark with any mass.

We study the dressed Polyakov loop in the framework of three-flavor NJL model:
\begin{equation}
  {\cal{L}} =  \bar{\psi}(i\gamma^{\mu}\partial_{\mu}-m)\psi
    + G_s \sum_a \Big\{ (\bar{\psi}\tau_a\psi)^2 + (\bar{\psi}i\gamma_5\tau_a\psi)^2
    \Big\}
    -  K \Big\{ {\rm Det}_f[\bar{\psi}(1+\gamma_5)\psi] + {\rm Det}_f[\bar{\psi}(1-\gamma_5)\psi]
   \Big\}.
\label{Lagr-flavor-3}
\end{equation}
Where $\psi=(u,d,s)^T$ denotes the transpose of the quark field, and
$m={\rm Diag}(m_u,m_d,m_s)$ is the corresponding mass matrix in the
flavor space. $\tau_a$ with $a=1,\cdots,N_f^2-1$ are the eight
Gell-Mann matrices, and ${\rm Det}_f$ means determinant in flavor
space. The last term is the standard form of the 't Hooft
interaction, which is invariant under $SU(3)_L\times SU(3)_R\times
U(1)_B$ symmetry, but breaks down the $U_A(1)$ symmetry.

\begin{figure}[h]
\begin{minipage}[t]{0.45\linewidth}
\centering
\includegraphics[width=7cm]{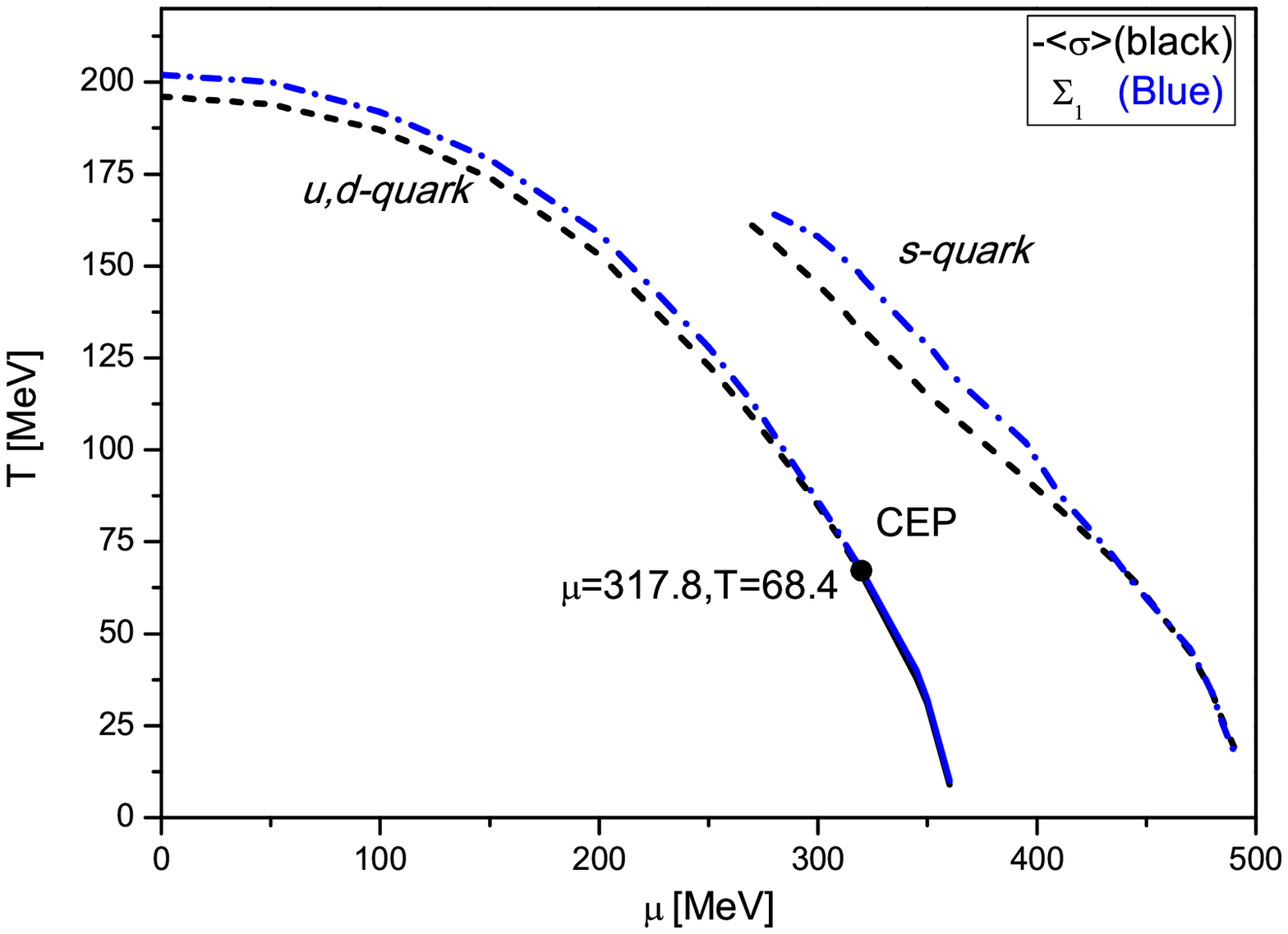}
\caption{Three-flavor phase diagram in the $T-\mu$ plane for the
case of $m_u=m_d=5{\rm MeV}$ and $m_s=140.7{\rm MeV}$.  \label{fig-SU3}}
\end{minipage}
\hfill
\begin{minipage}[t]{0.45\linewidth}
\centering
\includegraphics[width=7cm]{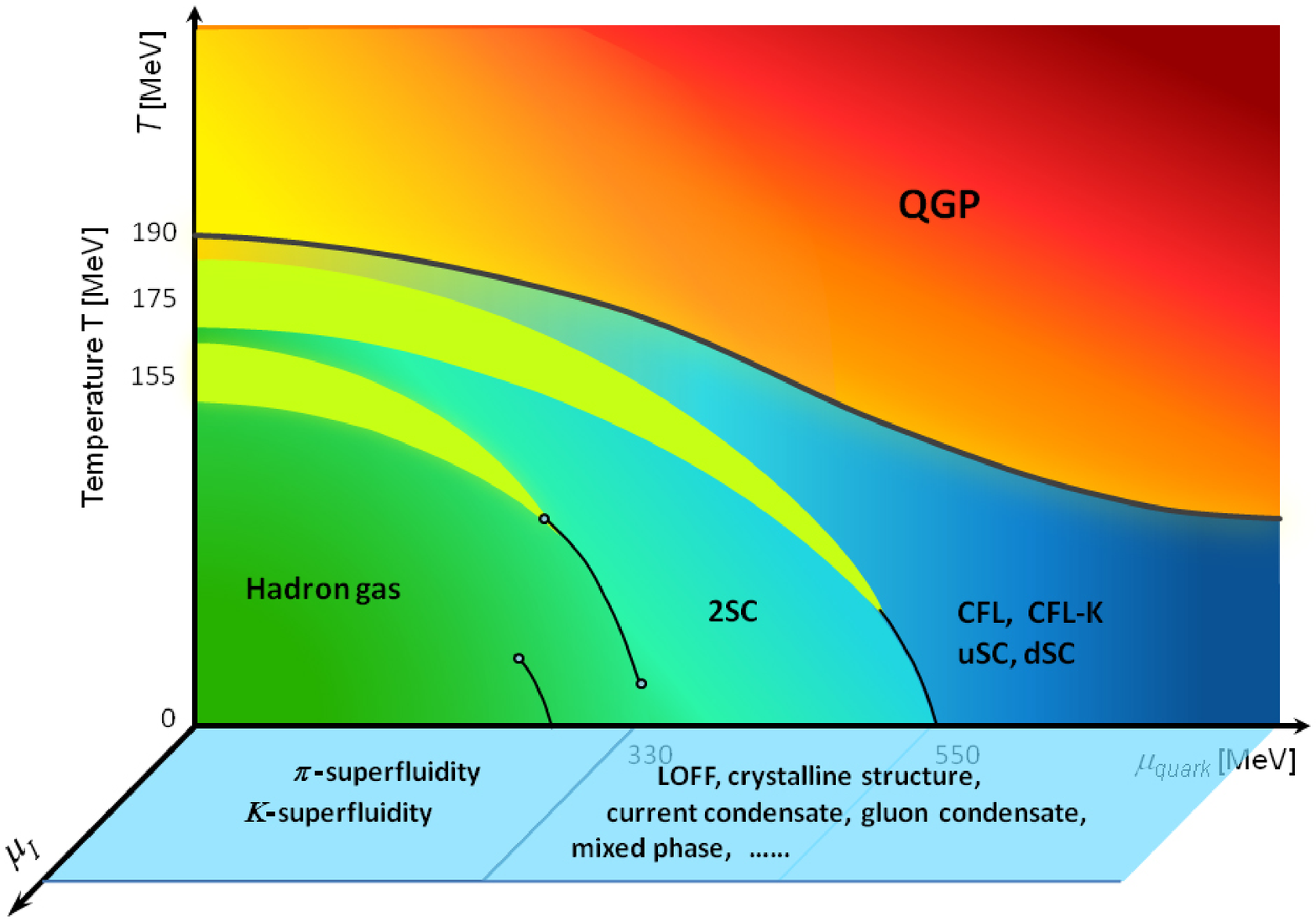}
\caption{Conjectured 3D QCD phase diagram at finite temperature $T$,
quark chemical potential $\mu_q$ and isospin chemical potential
$\mu_I$. \label{fig-conjecture-phase}}
\end{minipage}
\end{figure}

The chiral phase transition characterized by conventional chiral condensate
and the deconfinement phase transition characterized by the dressed Polyakov loop
$\Sigma_1$ are investigated in Ref. \cite{NJL-DPL}. For the two flavor case, our results
agree with that in the Dyson-Schwinger Equations \cite{DSE-DPL}.
The three-flavor phase diagram in the
$T-\mu$ plane for the case of $m_u=m_d=5{\rm MeV}$ and $m_s=140.7{\rm MeV}$
is shown in Fig.\ref{fig-SU3}.  It is found that the phase transitions are flavor
dependent, and there is a phase transition range for each flavor.
The transition range of $s$ quark is located at higher temperature
and higher baryon density than that of $u,d$ quarks. At low baryon
density region, it is found that the transition range of $u,d$
quarks are not separated too much from that of the $s$ quark,
however, the separation of the transition ranges for $u,d$ quarks
and $s$ quark become wider and wider with the increase of the
chemical potential. Based on above reulsts, in Fig. \ref{fig-conjecture-phase},
we show
our conjectured 3 dimension (3D) QCD phase diagram for finite
temperature $T$, quark chemical potential $\mu_q$ and isospin
chemical potential $\mu_I$.

\section{Gluodynamic model for color electric deconfinement and color magnetic confinement}
\label{sec-gluodynamic}

In the framework of QCD effective models, there is still no
dynamical model which can describe the chiral symmetry breaking and
confinement simultaneously. The main difficulty of effective QCD
model to include confinement mechanism lies in that it is difficult
to calculate the Polyakov loop analytically. Currently, the popular
models used to investigate the chiral and deconfinement phase
transitions are the Polyakov Nambu-Jona-Lasinio model (PNJL)
and Polyakov linear sigma model (PLSM) \cite{PNJL-PLSM},
where the Polyakov loop is introduce in the framework statistically.
A dynamical model for describing deconfinement phase transition is
still missing. In \cite{Xu:2011ud} , we introduce a pure gluondynamic
model with dimension-2 gluon condensate, and investigate how good
this model can capture the main feature of deconfinement phase transition.
In last decade, there have been growing interests in dimension-2 gluon condensates
$< g^2 A^2> $ in SU$(N_c)$ gauge theory \cite{Zakharov, D2-GC}, which is regarded to
have close relation with confinement.

The pure gluon part of QCD Lagrangian is described by
$\mr{L}_{G} = -\frac{1}{4}G_{\mu\nu}^aG^a_{\mu\nu}$ with
$G_{\m\n}^a = \p_\m{A_\n^a}-\p_\n{A_\m^a}+gf^{abc}{A_\m^b}{A_\n^c}$.
The gluon field can be decomposed into a
condensate field $\mb{A}_\m^a$ and a fluctuating field
$\mathscr{A}_{\mu}^a$ as,
$ A_\mu^a(x):=\mathbb{A}_{\mu}^a+\mathscr{A}_{\mu}^a (x)$
\cite {Celenza:1986th}.
Then the Lagrangian after this background expansion becomes
$\lag\mr{L}\rag_{\hat{\eta}}
=-\frac{1}{4}\left[\mr{GG} + 2m_g^2\mr{A}^2 + 4b\phi_0^4\right]$,
with $ m_g^2=\frac{9}{32}g^2\phi_0^2$ and  $b=\frac{9}{136}g^2$.
The gluon gets mass because of
the existence of nonperturbative dimension-2 gluon condensate.

\begin{figure}[h]
\begin{minipage}[t]{0.45\linewidth}
\centering
\includegraphics[width=7cm]{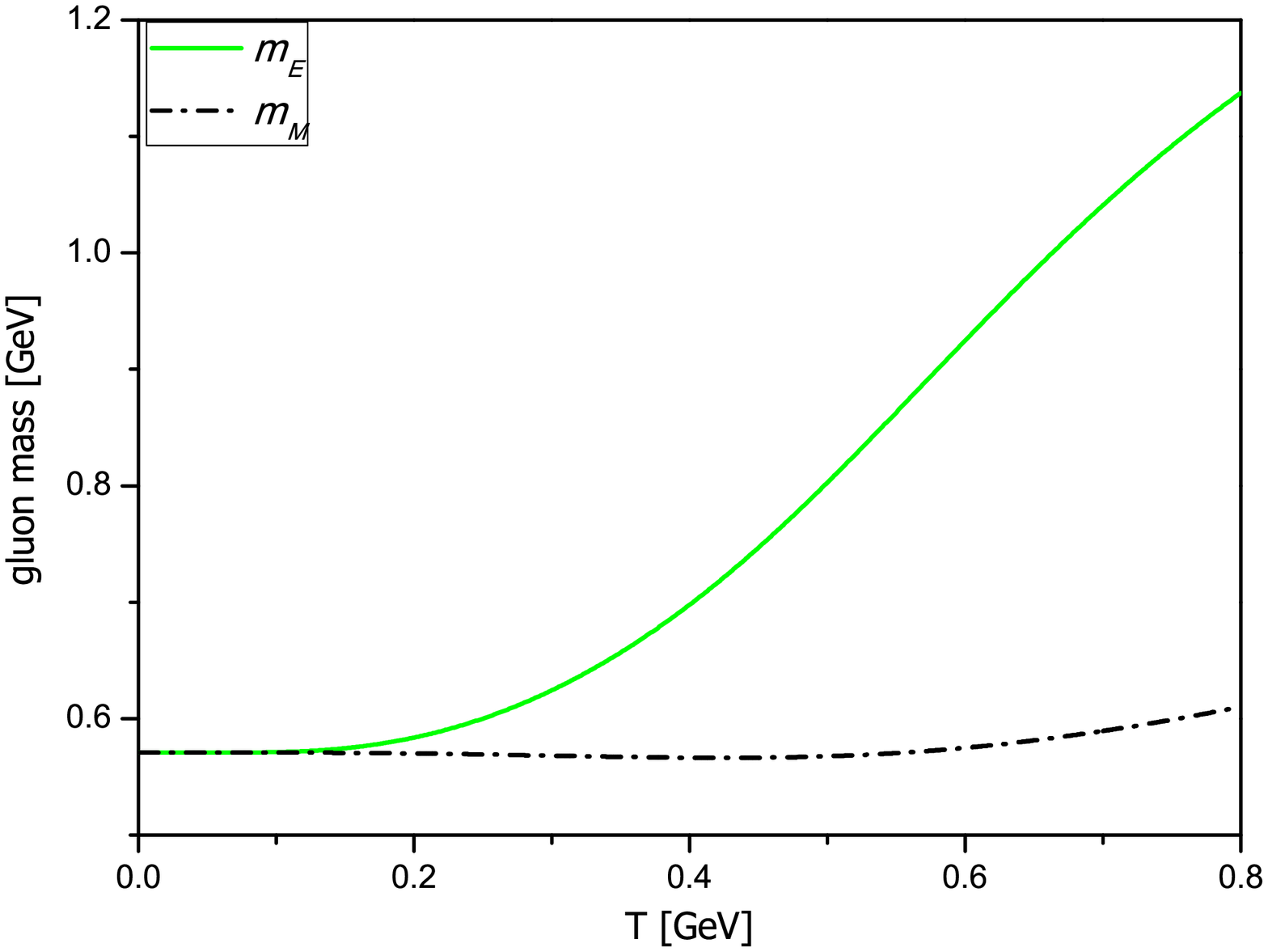}
\caption{The electric
and magnetic screening masses as functions of the temperature. \label{fig:mgT-lam03-xi8}}
\end{minipage}
\hfill
\begin{minipage}[t]{0.45\linewidth}
\centering
\includegraphics[width=7cm]{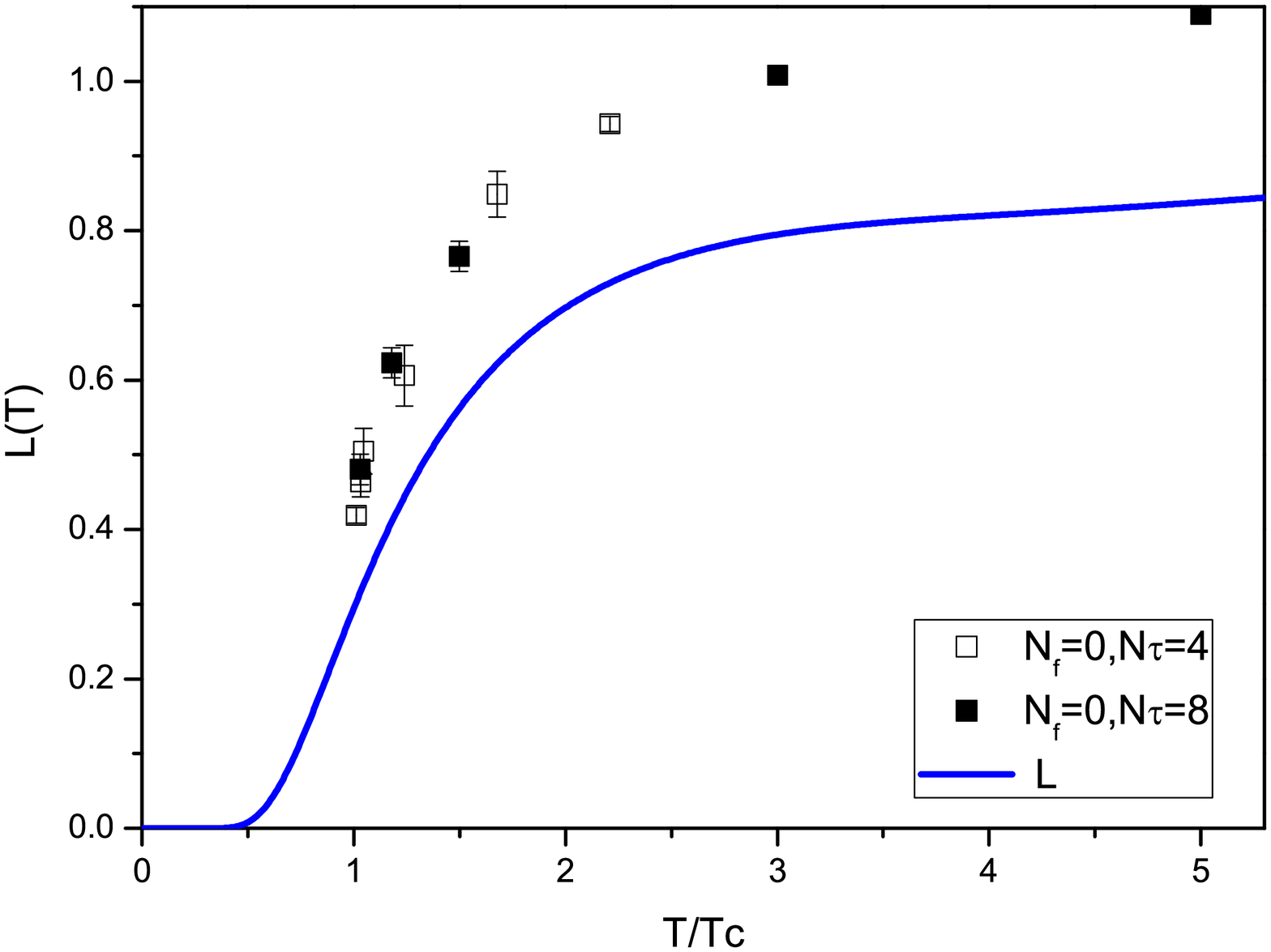}
\caption{The Polyakov loop expectation value as a function of
$T/T_c$ comparing with lattice result in
Ref.\cite{Kaczmarek:2002mc}. \label{fig:L-T}}
\end{minipage}
\end{figure}

At finite temperature, the electric and magnetic screening masses as functions of the
temperature are shown in Fig.\ref{fig:mgT-lam03-xi8}.  It is found that
the electric and magnetic components are degenerate at low temperature and
start to split at higher temperature, and then the electric
screening mass rise rapidly with $T$. Correspondingly, the Polyakov loop
expectation value as a function of $T/T_c$
in Fig. \ref{fig:L-T} is compared with lattice data in
Ref.\cite{Kaczmarek:2002mc}. It is found that the Polyakov loop
expectation value is zero in the vacuum and low temperature region,
then rise sharply  at high temperature. However, the magnetic screening mass of the gluons remains almost
the same as its vacuum value, which characterize the color magnetic confinement
feature of QCD.

\vskip 1cm \noindent

{\bf Acknowledgments}:  The work of M.H. is supported by CAS program
"Outstanding young scientists abroad brought-in", CAS key project KJCX2-EW-N01,
NSFC under the number of 10735040 and 10875134, and K.C.Wong Education Foundation,
Hong Kong.

\end{document}